%% file: cluster-k.tex
\documentclass[11pt]{article}
\usepackage{xspace}
\usepackage{graphics}
\usepackage{tabularx}
\usepackage{epsfig}
\usepackage{times}
\def\Box{\rule{2mm}{2mm}}

\newtheorem{fact}{Fact}[section]
\newtheorem{lemma}[fact]{Lemma}
\newtheorem{theorem}[fact]{Theorem}
\newtheorem{defn}[fact]{Definition}
\newtheorem{corollary}[fact]{Corollary}

\newtheorem{claim}[fact]{Claim}
\newcommand{\junk}[1]{}

\newcommand{\proof}[1]{{\noindent {\it Proof.} {#1} \Box \vskip \belowdisplayskip}}

\def\polylog{{\mathop{\mathrm{polylog}}\nolimits}}
\def\poly{{\mathop{\mathrm{poly}}\nolimits}}

\newcommand{\Max}{{\mbox{Max}}}

\newcommand{\eps}{\varepsilon}

\newcommand{\cluster}{{\tt Cluster}}
\newcommand{\cl}[1]{{T_{#1}}}

\newcommand{\eat}[1]{}
\newcommand{\Omegat}[1]{{{\widetilde \Omega} \left( #1 \right) }}
\title{Clustering with Spectral Norm and the $k$-means Algorithm}
\author{Amit Kumar \thanks{This work was done while the author was visiting 
Microsoft Research India Lab.} \\
 Dept. of Computer Science and Engg. \\
IIT Delhi, New Delhi  \\
email : {\tt amitk@cse.iitd.ac.in}
\and Ravindran Kannan \\
Microsoft Research India Lab. \\
Bangalore \\
email : {\tt kannan@microsoft.com}
}

\newcounter{myfootertablecounter}

\begin{document}
\begin{titlepage}
%\date{}
\maketitle
\thispagestyle{empty}
\begin{abstract}
 There has been much progress on efficient algorithms for clustering
data points generated by a mixture of $k$ probability distributions
under the assumption that the means of the distributions are well-separated,
i.e., the distance between the
means of any two distributions is at least $\Omega(k)$ standard
deviations. These results generally make heavy use of the generative model and
particular
properties of the distributions. In this paper, we show that a simple
clustering algorithm works without assuming any
generative (probabilistic) model. Our only assumption
is what we call a ``proximity condition'': the
projection of any data point onto the line joining its cluster center to any other
cluster center
is $\Omega(k)$ standard deviations closer to its own center than the other center.
Here the notion of standard deviations is based on the spectral norm of the matrix
whose rows represent the
difference between a point and the mean of the cluster to which it belongs. We show
that in
the generative models studied, our proximity condition is satisfied and so we are
able to derive
most known results for generative models as corollaries of our main result.
We also prove some new results for generative models - e.g.,  we can cluster all
but a small
fraction of points only assuming a bound on the variance.
Our algorithm relies on the well known $k$-means algorithm, and along the way, we prove
a result of independent interest --
that the $k$-means
algorithm converges to the ``true centers'' even in the presence of spurious points
provided the
initial (estimated) centers are close enough to the corresponding actual centers and
all but a small
fraction of the points satisfy the proximity condition. Finally, we present a new
technique for
boosting the ratio of inter-center separation to standard deviation. This allows us to 
prove results for learning mixture of a class of distributions under weaker separation 
conditions. 
\end{abstract}
\end{titlepage}

\section{Introduction}
Clustering is in general a hard problem.
%have applications  in diverse areas like computer vision,
%medical imaging, data mining, and machine learning (see e.g., \cite{lindsay}).
But, there has been a lot of research (see Section~\ref{sec:previous} for references) on proving that if we have data points generated by a mixture of $k$
probability distributions, then one can cluster the data points into the $k$ clusters, one corresponding to each
component, provided the means of the different components are well-separated. There are different notions of well-separated, but mainly, the
(best known) results can be qualitatively stated as:
\vspace*{-0.1in}
\begin{center}
``If the means of every pair of densities are at least $\poly(k)$ times standard deviations apart, then we can learn the mixture in polynomial time.''
\end{center}
\vspace*{-0.1 in}
\noindent
These results generally make heavy use of the generative model and particular properties of the distributions (Indeed, many of them specialize to Gaussians or independent Bernoulli trials).
In this paper, we make no assumptions on the generative model of the data. We are still able to derive  essentially the same result (loosely stated for now as):
\vspace*{-0.1 in}
\begin{center}
``If the projection of any data point onto the line joining its cluster center to any other cluster center is $\Omega(k)$ times
standard deviations closer to its own center than the other center (we call this the ``proximity condition''), then
we can cluster correctly in polynomial time.''
\end{center}
\vspace*{-0.1 in}
\noindent
First, if the $n$ points to be clustered form the rows of an $n\times d$ matrix $A$ and $C$ is the corresponding matrix of cluster centers (so each row of $C$ is one of $k$ vectors, namely the centers of $k$ clusters)
then note that the maximum directional variance (no probabilities here, the variance is  just the average squared distance from the center)
of the data in any direction is just
$$\frac{1}{n} \cdot \Max_{v:|v|=1} |(A-C) \cdot v|^2 = \frac{||A-C||^2}{n},$$
where $||A-C||$ is the spectral norm. So, spectral norm scaled by $1/\sqrt n$ will play the role of standard deviation in the above assertion. To our knowledge, this is the first result proving that clustering can be done in polynomial time in a general situation with only deterministic assumptions. It settles an open question raised in \cite{KannanV09}.

We will show that in the generative models studied, our proximity condition is satisfied and so we are able to derive all known results for generative models as corollaries of our theorem (with one qualification: whereas our separation is in terms of the whole data variance, often, in the case of Gaussians, one can make do with separations depending only on individual densities' variances -- see Section~\ref{sec:previous}.)

Besides Gaussians, the planted partition model (defined later) has also been studied; both these distributions have very ``thin tails'' and a lot of independence, so one can appeal to concentration results. In section~\ref{sec:boundedvariance}, we give a
clustering algorithm for a mixture of general densities for which we only assume bounds on the variance (and no further concentration). Based on our algorithm, we show how to classify all but an $\eps$ fraction of points in this model. Section~\ref{sec:previous}
 has references to recent work dealing with distributions which may not even have variance, but these results are only for the special class of product
densities, with additional constraints.

One crucial technical result we prove (Theorem~\ref{thm:key}) may be of independent interest. It shows that the good old $k-$means algorithm~\cite{lloyd}
converges to the ``true centers'' even in the presence of spurious points provided the
initial (estimated) centers are close enough to the corresponding actual centers
and all but an $\eps$ fraction of the points satisfy the proximity condition. Convergence (or lack of it) of the $k-$means algorithm is
again well-studied (\cite{OstrovskyRSS06, ArthurV06, Dasgupta03a, Har-PeledS05}). The result of~\cite{OstrovskyRSS06} (one of the few to formulate sufficient conditions for the
 $k-$means algorithm to provably work)
assumes the condition that  the optimal clustering with $k$ centers is substantially better than that  with fewer centers and shows that one iteration of $k-$means yields a near-optimal solution.
%and proves convergence.
We show in section~\ref{sec:convergence} that their condition implies proximity for all but an $\eps$ fraction of the points. This allows us to prove that 
our algorithm, which is again based on the $k-$means algorithm, gives a PTAS.

The proof of Theorem~\ref{thm:key} is based on Theorem
~\ref{thm:iterate} which shows that if current centers are close to the true centers, then misclassified points (whose nearest current center
 is not the one closest to the true center) are far away from true centers and so there cannot be too many of them. This is based on a clean
geometric argument shown pictorially in Figure~\ref{fig:classify}.
Our main theorem in addition allows for an $\eps$ fraction of ``spurious'' points which do
 not satisfy the proximity condition. Such errors have often proved difficult to account for.

As indicated, all results on generative models assume a lower bound on the inter-center separation in terms of the spectral norm.
In section~\ref{boosting}, we describe a construction (when data is from a generative model -- a mixture of distributions)
which boosts the ratio of inter-center separation to spectral norm.
The construction is the following: we pick two sets of samples $A_1,A_2,\ldots A_n$ and $B_1,B_2,\ldots B_n$ independently from the mixture. We define new points $X_1,X_2,\ldots X_n$, where $X_i$ is defined as $(A_i'\cdot B_1',A_i'\cdot B_2',\ldots A_i'\cdot B_n')$, where $'$ denotes that we have subtracted the mean (of the mixture.) Using this, we are able to reduce the dependence of inter-center separation on the minimum weight of
a component in the mixture that all models generally need. This technique of boosting is likely to have other applications.
\vspace*{-0.1 in}
\section{Preliminaries and the Main Theorem}
For a matrix $A$, we shall use $||A||$ to denote its spectral norm. For a vector $v$, we use $|v|$ to denote its length.
We are given $n$ points in $\Re^d$ which are divided into $k$ clusters -- $\cl{1}, \cl{2}, \ldots, \cl{k}$. Let $\mu_r$ denote the mean
of cluster $\cl{r}$ and $n_r$ denote $|T_r|$. Let $A$ be the $n \times d$ matrix with rows corresponding to the points.
% The set of points are from the rows of  an $m \times n$ matrix $A$.
Let $C$ be the $n \times d$ matrix where $C_i =
\mu_r$, for all $i \in \cl{r}$. We shall use $A_i$ to denote the $i^{th}$ row of $A$.
% Let $m_0$ denote the smallest
%size of a cluster.
%We shall also
%assume that there is a constant $\alpha_0$ such that $m_0 \geq \alpha_0 n$.
Let $$  \Delta_{rs} = \left( \frac{ck}{\sqrt{n_r}}+\frac{ck}{\sqrt{n_s}}\right)||A-C||,$$
where $c$ is a large enough constant.

\begin{defn}
  We say a point $A_i \in T_r$ satisfies the {\it proximity condition}
if for any $s\not= r$, the projection of $A_i$ onto the $\mu_r$ to $\mu_s$ line
is at least $\Delta_{rs}$ closer
to $\mu_r$ than to $\mu_s$. We let $G$ (for good) be the set of points satisfying the
proximity condition.
\end{defn}

Note that the proximity condition implies that  the distance between $\mu_r$ and $\mu_s$ must be at least $\Delta_{rs}$.
 We are now ready to state the theorem.

%% OLDER version of the main theorem
%%\begin{theorem}
%%\label{thm:main}
%% Let $F^{r,s}$ denote the projection on the line joining the means $\mu_r$ and $\mu_s$. Suppose the following separation condition
%%holds for all $r \neq s$ and all $i \in T_r$ :
%%\begin{eqnarray}
%%\label{eqn:condition}
%% ||F^{r,s}(A_{(i)}) - \mu_r)|| \leq ||F^{r,s}(A_{(i)}) - \mu_s|| - C \cdot \sqrt{\frac{k}{\alpha_0}} \cdot  \frac{||A-C||}{\sqrt{n}},
%%\end{eqnarray}
%%where $C$ is a large enough constant. Then we can correctly classify all the points.
%%\end{theorem}

%Let $D(r,s)$ denote $||\mu_r - \mu_s||$. The separation condition implies that %$D(r,s) \geq \frac{\beta \cdot ||A-C||}{\sqrt{n}}$.
%We shall assume without loss of generality the origin is the mean of the points.
\begin{theorem}\label{thm:main}
If $|G| \geq (1-\eps) \cdot n$, then we can correctly classify all but  $O(k^2  \eps \cdot n)$ points in polynomial time. In particular, if $\eps=0$, all points are classified correctly.
\end{theorem}

Often, when applying this theorem to learning a mixture of distributions, $A$ will correspond to a set of $n$ independent samples from the
mixture. We will denote the corresponding distributions by $F_1, \ldots, F_k$, and their relative weights by $w_1, \ldots, w_k$. Often,
$\sigma_r$ will denote the maximum variance along any direction of the distribution $F_r$, and $\sigma$ will denote $\max_r \sigma_r$. We denote
the minimum mixing weight of a distribution as $w_{\min}$. 
\vspace*{-0.1in}
\section{Previous Work}
\label{sec:previous}
Learning mixture of distributions is one of the central problems in machine learning. % As mentioned  earlier,
%we are given $n$ samples from a mixture distribution $F_1, \ldots, F_k$ in $d$ dimensions with relative weights
%$w_1, \ldots, w_k$, and we
%would like to group the points from the same distribution in one cluster.
 There is vast amount of
literature on learning mixture of Gaussian distributions. One of the most popular methods for this is the well
known EM algorithm which maximizes the log likelihood function \cite{Dempster77maximumlikelihood}. However, there
are few results which demonstrate that it converges to the optima solution. Dasgupta \cite{dasgupta99} introduced the
problem of learning distributions under suitable {\em separation conditions}, i.e., we assume that the distance between the
means of the distributions in the mixture is large, and the goal is to recover the original clustering of points (perhaps with
some error).

We first summarize  known results for learning mixtures of Gaussian distributions under separation conditions. We ignore logarithmic
factors in separation condition. We also ignore the minimum number of samples required by the various algorithms -- they are
often bounded by a polynomial in the dimension and the mixing weights. Let $\sigma_r$
be the maximum variance of the Gaussian $F_r$ in any direction. Dasgupta \cite{dasgupta99} gave an algorithm based on random
projection to learn mixture of Gaussians provided mixing weights of all distributions are about the same, and
$|\mu_i - \mu_j|$ is $\Omega( (\sigma_i + \sigma_j) \cdot \sqrt{n}). $  Dasgupta and Schulman \cite{DasguptaS07} gave
an EM based algorithm provided $|\mu_i - \mu_j|$ is $\Omega( (\sigma_i + \sigma_j) \cdot n^{\frac{1}{4}}). $  Arora and Kannan \cite{AroraKannan01}
also gave a learning algorithm with similar separation conditions. Vempala and Wang \cite{VempalaW04} were the first to demonstrate the
effectiveness of spectral techniques. For spherical Gaussians, their algorithm worked with a much weaker separation condition of
$\Omega( (\sigma_i + \sigma_j) \cdot k^{\frac{1}{4}})$ between $\mu_i$ and $\mu_j$. Achlioptas
and McSherry \cite{AchlioptasM05} extended this to  arbitrary Gaussians  with separation between $\mu_i$ and $\mu_j$ being
at least $\Omegat{\left( k + \frac{1}{\sqrt{\min(w_i, w_j})} \right) \cdot (\sigma_i + \sigma_j)}.$
 Kannan et. al. \cite{KannanSV08} also gave an algorithm for arbitrary Gaussians with the
 corresponding separation being $\Omega \left(\frac{k^{\frac{3}{2}}}{w_{\min}^2} \cdot (\sigma_i + \sigma_j)\right)$.
Recently, Brubaker and Vempala \cite{BrubakerV08} gave a learning algorithm where the separation only depends on the variance perpendicular to
a hyperplane separating two Gaussians (the so called ``parallel pancakes problem'').

Much less is known about learning mixtures of heavy tailed distributions. Most of the known results assume that each distribution is a
product distribution, i.e., projection along co-ordinate axes are independent. Often, they also assume some {\em slope condition} on the
line joining any two means. These slope conditions typically say that the unit vector along such lines does not lie almost entirely along
very few coordinates. Such a condition is necessay because if the only difference between two distributions were a single coordinate,
then one would require much stronger separation conditions. Dasgupta et. al. \cite{DasguptaHKS05} considered the problem of learning
product distributions of heavy tailed distributions when each component distribution satisfied the following mild condition :
$P [ |X-\mu| \geq \alpha R] \leq \frac{1}{2 \alpha}$. Here $R$ is the  half-radius of the distribution (these distributions can have
unbounded variance). Their algorithm could classify at least $(1-\eps)$ fraction of the points provided the distance between any
two means is at least $ \Omega \left( \frac{\sigma \cdot k^{\frac{5}{2}}}{\eps^2} \right)$. Here $R$ is the maximum half-radius of the distributions along
any coordinate. Under even milder assumptions on the distributions and a slope condition, they could correctly classify all but $\eps$ fraction of the
points provided the corresponding separation was $\Omega \left( \sigma \cdot \sqrt{\frac{k}{\eps} } \right)$. Their algorithm, however, requires
exponential (in $d$ and $k$) amount of time. This problem was resolved by Chaudhuri and Rao \cite{ChaudhuriR08a}.
Dasgupta et. al. \cite{DasguptaHKM07}
considered the problem of classifying samples from a mixture of arbitrary distributions with bounded variance in any direction. They showed that
if the separation between the means is $\Omega \left( \sigma k \right)$ and a suitable slope condition holds, then all the samples can be
correctly classified. Their paper also gives a general method for bounding the spectral norm of a matrix when the rows are independent (and some
additional conditions hold). We will mention this condition formally in Section~\ref{sec:app} and make heavy use of it.

Finally, we discuss the {\em planted partition model} \cite{McSherry01}. In this model, an instance consists of  a set of $n$ points, and
there is an implicit partition of these $n$ points into $k$ groups. Further, there is an (unknown) $k \times k$  matrix of prababilities $P$.
We are given a graph $G$ on these $n$ points, where an edge between two vertices from groups $i$ and $j$ is present with probability $P_{ij}$.
 The goal is to recover the actual partition of the points (and hence, an approximation to the matrix $P$ as well). We can think of this as a
special case of learning mixture of $k$ distributions, where the distribution $F_r$ corresponding to the $r^{th}$ part is as follows :  $F_r$ is
a distribution over $\{0,1\}^n$, one coordinate corresponding to each vertex. The coordinate corresponding to vertex $u$ is set to 1 with probability $P_{ij}$, where $j$
denotes the group to which $u$ belongs. Note that the mean of $F_r$, $\mu_r$, is equal to the vector $(P_{r \psi(u)})_{u \in V}$, where $\psi(u)$ denotes the
group to which the vertex $u$ belongs.  McSherry\cite{McSherry01} showed that if the following separation condition is satisfied, then one
can recover the actual partition of the vertex set with probability at least $1 - \delta$ -- for all $r, s$, $r \neq s$
\begin{eqnarray}
\label{eqn:planted}
|\mu_r - \mu_s|^2 \geq c \cdot \sigma^2 \cdot k \cdot \left( \frac{1}{w_{\min}} + \log \frac{n}{\delta} \right),
\end{eqnarray}
where $c$ is a large constant,  $w_{\min}$ is such that every group has size at least $w_{\min} \cdot n$, and $\sigma^2$ denotes $\max_{i,j} P_{ij}$.

There is a rich body of work on the $k$-means problem and heuristic algorithms for this problem (see for example~\cite{KumarSS10, OstrovskyRSS06} and references therein).
One of the most widely used algorithms for this problem was given by Lloyd~\cite{lloyd}. In this algorithm, we start with an arbitrary set of
$k$ candidate centers. Each point is assigned to the closest candidate center -- this clusters the points into $k$ clusters. For each cluster,
we update the candidate center to the mean of the points in the cluster. This gives a new set of $k$ candidate centers. This process is repeated
till we get a local optimum. This algorithm may take superpolynomial time to converge \cite{ArthurV06}. However, there is
 a growing body of work on proving that this algorithm gives a good clustering in polynomial time
if the initial choice of centers is good ~\cite{ArthurV07, AggarwalDK09, OstrovskyRSS06}. Ostrovsky et. al.~\cite{OstrovskyRSS06}
showed that a modification of the Lloyd's algorithm gives a PTAS for the $k$-means problem if there is a sufficiently large
separation between the means.
 Our result also fits in this general theme -- the $k$-means algorithm on a choice
of centers obtained from a simple spectral algorithm classifies the point correctly.
\vspace*{-0.1 in}
\section{Our Contributions}
Our main contribution is to show that a set of points satisfying a deterministic proximity condition (based on spectral norm) can be
correctly classified (Theorem~\ref{thm:main}). The algorithm is described in Figure~\ref{fig:alg}. It has two main steps -- first find an
initial set of centers based on SVD, and then run the standard $k$-means algorithm with these initial centers as seeds. In Section~\ref{sec:proof},
we show that after each iteration of the $k$-means algorithm, the set of centers come exponentially close to the true centers. Although both
steps of our algorithm -- SVD and the $k$-means algorithm -- have been well studied, ours is the first result which shows that {\em combining}
the two leads to a provably good algorithm. In Section~\ref{sec:app}, we give several applications of Theorem~\ref{thm:main}. We have the
following results for learning mixture of distriutions (we ignore poly-logarithmic factors in the discussion below) :
\begin{itemize}
\item Arbitrary Gaussian Distributions with separation $\Omega \left(\frac{ \sigma k}{\sqrt{w_{\min}}} \right)$ : as mentioned above,
this matches known results~\cite{AchlioptasM05, KannanSV08} except for the fact that  the separation condition between two distributions
depends on the maximum standard deviation (as compared
to standard deviations of these distributions only).
\item Planted distribution model with separation $\Omega \left( \frac{k \sigma}{\sqrt{w_{\min}}} \right)$ : this matches the result of
McSherry~\cite{McSherry01} except for a $\sqrt{k}$ factor which we can also remove with a more careful analysis.
\item Distributions with bounded variance along any direction : we can classify all but an $\eps$ fraction of points if the separation
between means is at least $\Omega \left( \frac{k \sigma}{\sqrt{\eps}} \right)$. Although results are known for classifying (all but a small fraction) points
from mixtures of distributions with unbounded variance \cite{DasguptaHKS05, ChaudhuriR08a}, such results work for product distributions only.
\item PTAS using the  $k$-means algorithm   : We show that the separation condition of Ostrovsky et. al.~\cite{OstrovskyRSS06} is stronger than the proximity condition.  Using this
fact, we are also able to give a PTAS based on  the $k$-means algorithm. 
\end{itemize}
Further, ours is the first algorithm which applies to all of the above settings. In Section~\ref{boosting}, we give a general technique for working
with weaker separation conditions (for learning mixture of distributions). Under certain technical conditions described in Section~\ref{boosting},
we give a construction which increases the spectral norm of $A-C$ at a much faster rate than the increase in inter-mean distance as we
increase the number of samples. As applications of this technique, we have the following results :
\begin{itemize}
\item Arbitrary Gaussians with separation  $\Omega \left(  \sigma k \cdot \log \frac{d}{w_{\min}} \right)$ : this is the first result for arbitrary
Gaussians where the separation depends only logarithmically on the minimum mixture weight.
\item Power-law distributions with sufficiently large (but constant) exponent $\gamma$ (defined in equation~(\ref{eqn:heavytail})) :
We prove that we can learn all but $\eps$ fraction of samples provided the separation between means is $\Omega \left(
 \sigma k \cdot \left( \log \frac{d}{w_{\min}} + \frac{1}{\eps^{\frac{1}{\gamma}}} \right) \right)$. For large
values of $\gamma$, it significantly reduces the dependence on $\eps$.
\end{itemize}
We expect this technique to have more applications.

\vspace*{-0.1 in}
\section{Proof of Theorem~\ref{thm:main}}
\label{sec:proof}
Our algorithm for correctly classifying the points will run in several iterations. At the beginning of each iteration, it will have a set of
$k$ candidate points. By a Lloyd like step, it will replace these points by another set of $k$ points. This process will go on for polynomial
number of steps.

%%$k$
%%{\em candidate points} $\mu^{(\ell)}_1, \ldots, \mu^{(\ell)}_k$. These points will
%% satisfy the following invariant for $r = 1, \ldots, k$ :  $$ || \mu_r - \mu^{(\ell)}_r || \leq \gamma
%%\cdot \sqrt{\frac{k}{\alpha_0}} \cdot \frac{||A-C||}{2^\ell \cdot \sqrt{n}},$$
%%where $\gamma$ is a large enough constant, but smaller than the constant $C$ used in condition~\ref{eqn:condition}.
 %The parameters $\alpha_0, \gamma$ and $\beta$ will obey  $\frac{k}{\alpha_0} \ll \gamma \ll \beta$ (where $\ll$ denotes ``much less'' informally).

\begin{figure}[htb]
  \vspace*{-0.1in}
  \hrule
  \begin{tabularx}{\textwidth}{|X|}
  \vspace*{-0.15 in}
\begin{enumerate}
    \item ({\bf Base case}) Let ${\hat A}_i$ denote the projection of the points on the best $k$-dimensional subspace found by computing SVD of $A$.
Let $\nu_r, r = 1, \ldots, k,$ denote the centers of a (near)-optimal solution to the $k$-means problem for the points ${\hat A}_i$.
    \item  For $\ell = 1,2, \ldots$ do
\begin{itemize}
\item[(i)]  Assign each point $A_i$ to the closest point among  $\nu_r, r=1, \ldots, k$. Let $S_r$ denote the
set of points assigned to $\nu_r$.
\item[(ii)] Define $\eta_r$ as the mean of the points $S_r$. Update $\eta_r$, $r = 1, \ldots, k$ as the new centers, i.e., set
$\nu_r = \eta_r$ for the next iteration.
\end{itemize}
\end{enumerate}
%\vspace*{-1in}
  \end{tabularx}
  \hrule
  \vspace*{-0.1in}
  \caption{Algorithm {\cluster}}
  \label{fig:alg}
\end{figure}

%%It is intuitively clear that once we prove this fact, we will be done. We will show that
%% picking $\ell$ to be large enough will imply Theorem~\ref{thm:main}.
The iterative procedure is described in Figure~\ref{fig:alg}. In the first step, we can use any constant factor approximation algorithm for
the $k$-means problem.
Note that the algorithm is same as Lloyd's algorithm, but we start with a special set of initial points as described in the algorithm.
%% We now prove that the invariant conditions hold for all $\ell$.
 We now prove that after the first step (the base case), the estimated centers are close to the actual ones -- this
case follows from \cite{KannanV09}, but we prove it below for sake of completeness.
\begin{lemma}
\label{lem:case0}
({\bf Base Case}) After  the first step of the algorithm above, $$ |\mu_r - \nu_r| \leq 20 \sqrt{k} \cdot \frac{ ||A-C||}{\sqrt{n_r}}. $$
\end{lemma}
\proof{ Suppose, for sake of contradiction, that there exists an $r$ such that all the centers $\nu_1, \ldots, \nu_k$ are at least
$\frac{20 \sqrt{k} \cdot ||A-C||}{ \sqrt{n_r}}$ distance away from $\mu_r$. Consider the points in $\cl{r}$. Suppose $A_i \in \cl{r}$ is
assigned to the center $\nu_{c(i)}$ in this solution.
 The assignment cost for these points in this optimal $k$-means solution
 is
\begin{eqnarray}
\nonumber
\sum_{i \in \cl{r}} |{\hat A}_i - \nu_{c(i)}|^2 & = & \sum_{i \in \cl{r}} |(\mu_r-\nu_{c(i)}) - (\mu_r-{\hat A}_i)|^2 \\
\label{eq:ab}
 & \geq &
\frac{|\cl{r}|}{2} \cdot \left( \frac{20 \sqrt{k} \cdot ||A-C||}{\sqrt{n_r}} \right)^2 -
\sum_{i \in \cl{r}} |\mu_r - {\hat A}_i|^2 \\
\label{eq:fto2}
& \geq & 20k \cdot ||A-C||^2 - 5k \cdot ||A-C||^2  \ = \ 15k ||A-C||^2
\end{eqnarray}
where inequality~(\ref{eq:ab}) follows from the fact that for any two numbers $a, b$, $(a-b)^2 \geq \frac{a^2}{2} - b^2$; and    inequality~(\ref{eq:fto2}) follows from the fact that $||{\hat A} - C||_F^2 \leq 5k \cdot ||A-C||^2. $
But this is a contradiction, because one feasible solution to the $k$-means problem is to assign points in ${\hat A}_i, i \in \cl{s}$ to $\mu_s$ for
$s = 1, \ldots, k$ -- the cost of this solution is $ ||{\hat A}-C||_F^2 \leq 5k ||A-C||^2. $
}

Observe that the lemma above implies that there is a unique center $\nu_r$ associated with each $\mu_r$.
We now prove a useful lemma which states that removing small number of points from a cluster $\cl{r}$ can move the mean of the
remaining points by only a small distance.

\begin{lemma}
\label{lem:move}
Let $X$ be a subset of $\cl{r}$. Let $m(X)$ denote the mean of the points in $X$. Then
$$ | m(X) - \mu_r| \leq \frac{||A-C||}{\sqrt{|X|}}. $$
\end{lemma}
\proof{ Let $u$ be unit vector along $m(X) - \mu_r$. Now,
\begin{eqnarray*}
|(A-C) \cdot u | \geq \left( \sum_{i \in X} \left( (A_i-\mu_r) \cdot u \right)^2 \right)^{\frac{1}{2}} 
\  \geq  \frac{1}{\sqrt{|X|}} \left( \sum_{i \in X} |(A_i - \mu_r) \cdot u | \right) 
 \geq \ \sqrt{|X|} \cdot |m(X)-\mu_r|
\end{eqnarray*}
But, $|(A-C) \cdot u | \leq ||A-C||$. This proves the lemma.
}

\begin{corollary}
\label{cor:move}
Let $Y \subseteq \cl{s}$ such that $|\cl{s}-Y| \leq \delta \cdot n_s$, where $\delta < \frac{1}{2}$.
 Let $m(Y)$ denote the mean of the points in $Y$.
Then $$|m(Y)-\mu_s| \leq \frac{2 \cdot \sqrt{\delta} \cdot ||A-C||}{\sqrt{n_s}}. $$
\end{corollary}
\proof{ Let $X$ denote $\cl{s}-Y$. We know that $\mu_s \cdot |\cl{s}| = |X| \cdot m(X) + |Y| \cdot m(Y). $
So we get
\begin{eqnarray*}
|m(Y) - \mu_s | \  =  \  \frac{|X|}{|Y|} \cdot |m(X) - \mu_s | 
\ \leq  \  \frac{\sqrt{X}}{|Y|} \cdot ||A - C||
\end{eqnarray*}
where the inequality above follows from Lemma~\ref{lem:move}. The result now follows because $|Y| \geq \frac{n_s}{2}.$
}
Now we show that  if the estimated centers are close to the actual centers, then one iteration of the second step in the algorithm will reduce this separation
by at least half.

\noindent
{\bf Notation :}
\vspace*{-0.1in}
\begin{itemize}
\item $\nu_1,\nu_2,\ldots \nu_k$ denote the current centers at the beginning of an iteration in the second step of the algorithm, where $\nu_r$ is the current center closest to $\mu_r$.
\item $S_r$ denotes the set of points $A_i$ for which the closest current center is $\nu_r$.
\item $\eta_r$ denotes the mean of points in $S_r$; so $\eta_r$ are the new centers. Let $\delta_r = |\mu_r - \nu_r|$.
\end{itemize}
The theorem below shows that the set of misclassified points (which really belong to $T_r$, but have $\nu_s, s \neq r$,
 as the closest current center) are not too many in number. The proof first shows that any misclassified point must be far away from $\mu_r$ and since the sum of squared distances from $\mu_r$ for all points in $T_r$ is bounded, there cannot be too many.
\begin{theorem}
\label{thm:iterate}
Assume that  $\delta_r+\delta_s\leq \Delta_{rs}/16$ for all $r\not= s$. Then,
\begin{eqnarray}
\label{eqn:iterate2}
|T_r\cap S_s\cap G|\leq \frac{6ck \cdot ||A-C||^2(\delta_r^2+\delta_s^2)}{\Delta_{rs}^2|\mu_r-\mu_s|^2}
%%&\text{ for any } W\subseteq T_r\cap S_s\; :\; |m(W)-\mu_s|\leq \frac{100||A^r-C^r||}{\sqrt{|W|}} .\label{iterate-1}
\end{eqnarray}

Further, for any $W\subseteq T_r\cap S_s$,
\begin{eqnarray}
\label{eqn:iterate1}
|m(W)-\mu_s|\leq \frac{100 \cdot ||A-C||}{\sqrt{|W|}}
\end{eqnarray}

\end{theorem}

\proof{  Let $\bar v$ denote the projection of vector $v$ to the affine space $V$ spanned by $\mu_1,\ldots
\mu_k,\nu_1,\ldots, \nu_k$ and $\eta_1,\eta_2,\ldots \eta_k$.
Assume $A_i\in T_r\cap S_s\cap G$.
Splitting $\bar A_i$ into its projection along the line $\mu_r$ to $\mu_s$ and
the component orthogonal to it, we can write
$$\bar A_i= \frac{1}{2}(\mu_r+\mu_s) + \lambda (\mu_r-\mu_s) +u,$$
where $u$ is orthogonal to $\mu_r-\mu_s$. Since $\bar A_i$
is closer to $\nu_s$ than to $\nu_r$, we have
\begin{eqnarray*}
&{\bar A_i} \cdot (\nu_s-\nu_r)\geq \frac{1}{2}(\nu_s-\nu_r)\cdot (\nu_r+\nu_s)\\
\mbox{i.e.,} &\frac{1}{2}(\mu_r+\mu_s) \cdot (\nu_s-\nu_r)+\lambda (\mu_r-\mu_s)  \cdot (\nu_s-\nu_r)+u\cdot (\nu_s-\nu_r)
\geq \frac{1}{2}(\nu_s-\nu_r) \cdot (\nu_s+\nu_r).
\end{eqnarray*}

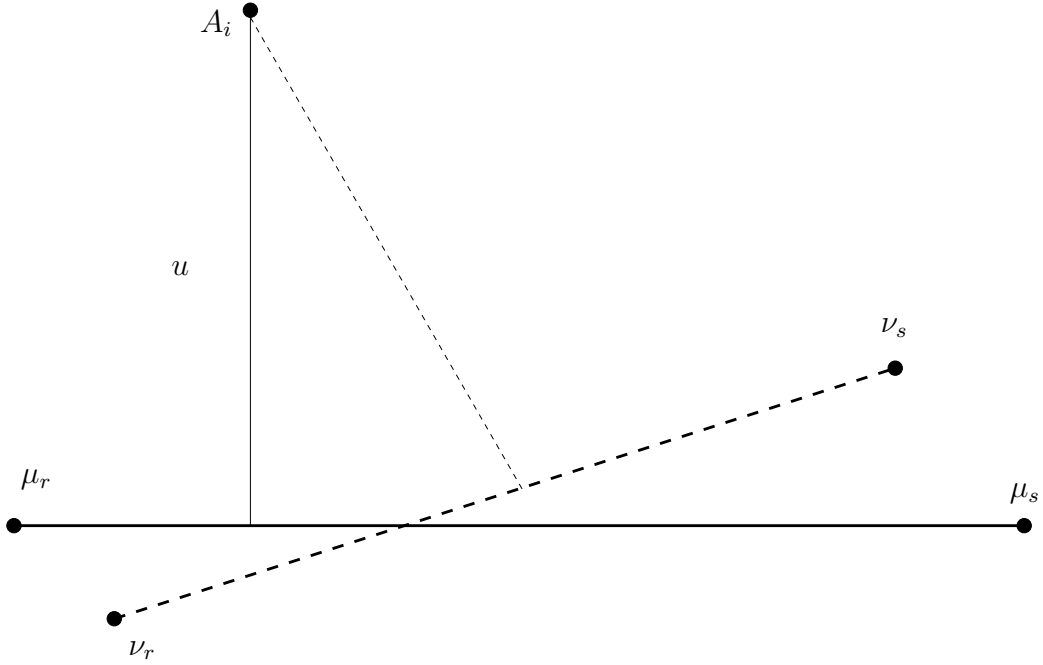
\begin{figure}
\begin{center}
  % Requires \usepackage{graphicx}
  %\includegraphics[width=5in]{Doc1.ps}\\
  % \epsfig{file=classify.eps, width=5in}
\input{classify.pstex_t}
\caption{Misclassified $A_i$}\label{fig:classify}
\end{center}
\end{figure}

We have $u\cdot (\nu_s-\nu_r)=u\cdot ((\nu_s-\mu_s)-(\nu_r-\mu_r))$ since $u$ is orthogonal
to $\mu_r-\mu_s$. The last quantity is at most $|u|\delta$, where $\delta=\delta_r+\delta_s$. Substituting this we get
\begin{eqnarray}
\nonumber
& \frac{1}{2} \cdot (\mu_r + \mu_s  - \nu_r - \nu_s) \cdot (\nu_s - \nu_r) + \lambda(\mu_r - \mu_s) \cdot (\nu_s - \nu_r) + |u| \cdot \delta \geq 0 \\
%\nonumber
%\mbox{i.e., } & \frac{1}{2} \cdot |\delta| \cdot (\delta + |\mu_r - \mu_s|) - \lambda |\mu_r - \mu_s|^2 + \lambda \delta |\mu_r - \mu_s| + |u| \cdot
%\delta \geq 0 \\
\label{u-length}
\mbox{i.e., }&\frac{\delta^2}{2}+\frac{\delta}{2}|\mu_r-\mu_s| -\lambda |\mu_r-\mu_s|^2+\lambda \delta |\mu_r-\mu_s|
+|u|\delta\geq 0.
\end{eqnarray}
Now,
\begin{eqnarray*}
|\bar A_i-\mu_r| & = & \left| \left( \frac{1}{2} - \lambda \right) \cdot (\mu_s - \mu_r) + u \right| \\
& \geq & |u| 
\  \geq \  \frac{\lambda}{\delta} \cdot |\mu_r - \mu_s|^2 - \lambda |\mu_r - \mu_s| - \frac{\delta}{2} - \frac{|\mu_r - \mu_s|}{2}  \ \ \ \ \ \ \
\mbox{using (\ref{u-length})} \\
& \geq & \frac{\Delta_{rs}|\mu_r-\mu_s|}{64\delta},
\end{eqnarray*}
where the last inequality follows from the fact that
 $\lambda \geq \frac{\Delta_{rs}}{2|\mu_r-\mu_s|}$ (proximity condition) and the assumption that $\delta \leq \Delta_{rs}/16$.
Therefore,  we have
$$ |T_r\cap S_s\cap G| \cdot \frac{\Delta_{rs}^2|\mu_r-\mu_s|^2}{c\delta^2}\leq \sum_{i\in T_r\cap S_s\cap G}|\bar A_i-\mu_r|^2
\leq \sum_{i\in T_r}|\bar A_i-C_i|^2.$$
If we take a basis $u_1,u_2,\ldots u_p$
of $V$, we see that $\sum_{i\in T_r}|\bar A_i - C_i|^2=\sum_{t=1}^p \sum_{i\in T_r}|(\bar A_i-C_i) \cdot u_t|^2
=\sum_{t=1}^p ||A-C||^2\leq 3k ||A-C||^2$, which proves the first statement of the theorem.

For the second statement, we can write $m(W)$ as
\begin{eqnarray*}
m(W)&= \frac{1}{2}(\mu_r+\mu_s) + \lambda (\mu_r-\mu_s) +u,
\end{eqnarray*}
where, $u$ is orthogonal to $\mu_r - \mu_s$. Since $m(W)$
is the average of points in $S_s$, we get (arguing as for (\ref{u-length})):
$$|u|\geq \frac{\lambda}{10\delta}|\mu_r-\mu_s|^2.$$
Now, we have
\begin{eqnarray*}
|m(W)-\mu_r|^2 = |u|^2+\left( \lambda-\frac{1}{2} \right)^2|\mu_r-\mu_s|^2, \ \mbox{and} \
|m(W)-\mu_s|^2 = |u|^2+\left(\lambda+\frac{1}{2} \right)^2|\mu_r-\mu_s|^2
\end{eqnarray*}
If $\lambda \leq 1/4$, then clearly, $|m(W)-\mu_s|\leq 4|m(W)-\mu_r|$. If
$\lambda>1/4$, then we have
$|u|\geq \frac{\lambda}{10\delta}|\mu_r-\mu_s|^2\geq \frac{1}{2} \cdot \left( \lambda +\frac{1}{2} \right) \cdot |\mu_r-\mu_s|$
because $\frac{|\mu_r - \mu_s|}{\delta} \geq 16$.
This again yields $|m(W)-\mu_s|\leq 4|u|\leq 4|m(W)-\mu_r|$.
Now, by Lemma~\ref{lem:move}, we have $|m(W)-\mu_r|\leq \frac{||A-C||}{\sqrt{|W|}}$,
so the second statement in the theorem. }

We are now ready to prove the main theorem of this section which will directly imply Theorem~\ref{thm:main}. This shows that $k-$means converges if
the starting centers  are close enough to the corresponding true centers.
 To gain intuition, it is best to look at the case $\eps=0$, when all points satisfy the proximity condition. Then the theorem says that if $|\nu_s-\mu_s|\leq \frac{\gamma||A-C||}{\sqrt{n_s}}$ for all $s$, then $|\eta_s-\mu_s|\leq
\frac{\gamma||A-C||}{2\sqrt{n_s}}$, thus halving the upper bound of the distance to $\mu_s$ in each iteration.
\begin{theorem}
\label{thm:key}
%%Let $\delta$ denote $\max_r \delta_r$. Then $$ |\eta_s - \mu_s | \leq \max \left( \frac{\delta}{2}, \frac{400 \sqrt {k \eps n} ||A - C||}{{n_s}} \right). $$
If $$\delta_s \leq \max \left(  \frac{\gamma \cdot ||A-C||}{ \sqrt{n_s}}, 800 \sqrt {k \eps n} \cdot \frac{||A-C||}{n_s} \right), $$
for all $s$ and a  parameter $\gamma\leq ck/50$,
 then $$ |\eta_s - \mu_s| \leq \max \left(  \frac{\gamma \cdot ||A-C||}{ 2 \sqrt{n_s}}, 800 \sqrt {k \eps n} \cdot \frac{||A-C||}{n_s} \right), $$
for all $s$.
\end{theorem}
\proof{ Let
$n_{rs},\mu_{rs} $ denote the number and mean respectively of $T_r\cap S_s\cap G$ and $n'_{rs},\mu'_{rs}$
of $(T_r\setminus G)\cap S_s$. Similarly, define $n_{ss}$ and $\mu_{ss}$ as the size and mean of the points in $\cl{s} \cap S_s$.
We get
\begin{eqnarray*}
|S_s|\eta_s&=n_{ss}\mu_{ss}+\sum_{r\not= s}n_{rs}\mu_{rs}+\sum_r n'_{rs}\mu'_{rs}.
\end{eqnarray*}
We have
$$|\mu_{ss}-\mu_s|\leq \frac{\sqrt{|S_s|-n_{ss}}}{n_{ss}}||A-C||\; ;\; |\mu_{rs}-\mu_s|\leq \frac{100 ||A-C||}{\sqrt{n_{rs}}}\; ;\;
|\mu'_{rs}-\mu_s|\leq \frac{100 ||A-C||}{\sqrt{n'_{rs}}},$$
where the first one is from Corollary~\ref{cor:move} (it is easy to check from the first statement of Theorem~\ref{thm:iterate} that
$n_{ss} \geq n_s/2$)  and the last two are from the second statement in Theorem~\ref{thm:iterate}.

Now using the fact that length is a convex function, we have
\begin{eqnarray*}
|\eta_s-\mu_s|&\leq & \frac{n_{ss}}{|S_s|}|\mu_{ss}-\mu_s| +\sum_{r\not= s}\frac{n_{rs}}{|S_s|}|\mu_{rs}-\mu_s|+\sum_r \frac{n'_{rs}}{{|S_s|}}|\mu'_{rs}-\mu_s|\\
&\leq &  200 ||A-C||\left( \frac{\sqrt{ |S_s|-n_{ss}}}{n_s}+\sum_{r\not= s}\frac{\sqrt{n_{rs}}}{n_s}+\sum_r \frac{\sqrt{n'_{rs}}}{n_s}\right) \\
& \leq & 400 ||A-C||\left(\sum_{r\not= s}\frac{\sqrt{n_{rs}}}{n_s}+\sum_r \frac{\sqrt{n'_{rs}}}{n_s}\right)
\end{eqnarray*}
since $|S_s| - n_{ss} = \sum_{r \neq s} n_{rs} + \sum_s n_{rs}'$.
Let us look at each of the terms above. Note that $n_{rs} \leq \frac{24ck ||A-C||^2 \max(\delta_r, \delta_s)^2 }{\Delta_{rs}^2 |\mu_r - \mu_s|^2}$ (using
Theorem~\ref{thm:iterate}). So
\begin{eqnarray*}
\sum_{r \neq s} \frac{\sqrt{n_{rs}}}{n_s} & \leq & \frac{5 \sqrt{c k} ||A-C||}{n_s} \sum_{r \neq s} \frac{\max(\delta_r, \delta_s)}{\Delta_{rs} \cdot |\mu_r - \mu_s|} \\
%%& \leq & \frac{\delta \sqrt{12 c k} ||A-C||}{n_s} \sum_{r \neq s} \frac{n_s}{c^2 k^2 ||A-C||^2} \leq \frac{\delta}{c ||A - C||}
& \leq & \frac{5 \sqrt{c k} ||A-C||^2 }{n_s} \sum_{r \neq s} \frac{1}{\Delta_{rs} \cdot |\mu_r - \mu_s|} \cdot \left( \frac{\gamma}{\sqrt{\min(n_r, n_s)}}
+  \frac{800 \sqrt{k \eps n}}{{\min(n_r, n_s)}} \right) \\
& \leq & \frac{5 \sqrt{c k}}{n_s} \sum_{r \neq s} \frac{\min(n_r, n_s)}{c^2 k^2} \cdot \left( \frac{\gamma}{\sqrt{\min(n_r, n_s)}}
+  \frac{800 \sqrt{k \eps n}}{{\min(n_r, n_s)}} \right) \\
& \leq & \frac{\gamma}{c  \sqrt{n_s}} + \frac{\sqrt{k \eps n}}{{n_s}}
\end{eqnarray*}

Also, note that $\sum_r n_{rs}' \leq \eps n$. So we get
$ \sum_r \frac{\sqrt{n_{rs}'}}{n_s} \leq \frac{\sqrt{k \eps n}}{n_s}. $
Assuming $c$ to be large enough constant  proves the theorem.
}

Now we can easily finish the proof of Theorem~\ref{thm:main}. Observe that after the base case in the algorithm, the statement of Theorem~\ref{thm:key}
holds with $\gamma = 20 \sqrt{k}$. So after enough number of iterations of the second step in our algorithm, $\gamma$ will become very small and so we will
get $$\delta_s \leq 800 \sqrt {k \eps n} \cdot \frac{||A-C||}{n_s} , $$ for all $s$. Now substituting this is Theorem~\ref{thm:iterate}, we get
\begin{eqnarray*}
|T_r \cap S_s \cap G|  \leq
\frac{800^2 \cdot 24 ck \cdot ||A-C||^4 \cdot \eps \cdot n \cdot k}{\Delta_{rs}^2 |\mu_r - \mu_s|^2 \cdot \min(n_r, n_s)^2} \leq \eps n
\end{eqnarray*}
Summing over all pairs $r, s$ implies Theorem~\ref{thm:main}.
\vspace*{-0.1in}
\section{Applications}
\label{sec:app}
We now give applications of Theorem~\ref{thm:main} to various settings. One of the main technical steps here would be to bound the spectral
norm of a random $n \times d$ matrix
$Y$ whose rows are chosen independently. We use the following result from \cite{DasguptaHKM07}. Let $D$ denote the
matrix $E[Y^TY]$. Also assume that $n \geq d$.
% Let $Y$ denote the matrix $X-Z$, and $D$ denote the matrix $E[Y^TY]$ where the expecation is over the choice of
%$A$ and $B$. Then the following fact follows from \cite{DasguptaHKM07}.
\begin{fact}
\label{fact:nice}
Let $\gamma$ be such  that $\max_i |Y_i| \leq \gamma \sqrt{n}$ and $||D|| \leq \gamma^2 n$. Then $||Y|| \leq \gamma \cdot \sqrt{n} \cdot
\polylog(n)$ with high probability.
\end{fact}
\subsection{Learning in the planted distribution model}
In this section, we show that McSherry's result \cite{McSherry01} can be derived as a corollary of our main theorem.
Consider an instance of the planted distribution model which satisfies the  conditon~(\ref{eqn:planted}).
We would like to show that with high probability, the points satisfy the proximity condition.
Fix a point $A_i \in \cl{r}$. We will show that the probability that it does not satisfy this condition is
at most $\frac{\delta}{n}$. Using union bound, it will then follow
that the proximity condition is satisfied with probability at least $1 - \delta$.

Let $s \neq r$. Let $v$ denote the
unit vector along $\mu_r - \mu_s$. Let $L_{rs}$ denote the line joining $\mu_r$ and $\mu_s$,
 and $\hat {A_i}$ be the projection of $A_i$ on $L_{rs}$.
The following result shows that the distance between $\hat {A_i}$ and $\mu_r$ is small with high probability.
\begin{lemma}
Assume $\sigma \geq \frac{3\log n}{n}, $ where $\sigma = \max_{i,j} \sqrt{P_{ij}}$.
With probability at least $1 - \frac{\delta}{n \cdot k}$, $$ |{\hat {A_i}} - \mu_r| \leq ck \cdot \sigma \cdot \left( \log \left( \frac{n}{\delta} \right)
+ \frac{1}{\sqrt{w_{\min}}} \right), $$
 where $c$ is a large constant. % and $\sigma$ denotes $\max_{i,j} \sqrt{P_{ij}}$.
\end{lemma}
\proof{
For a vector $A_i$, we use $A_{ij}$ to denote the coordinate of $A_i$ at position $j$. Define $\mu_{rj}$ similarly.
First observe that $ |{\hat {A_i}} - \mu_r| = |v \cdot (A_i - \mu_r)|$.
The coordinates of $v$ corresponding to points belonging to a particular cluster are same -- let $v^{t}$ denote this value for cluster $\cl{t}$. So we get
\begin{eqnarray*}
|v \cdot (A_i - \mu_r)|  \leq \sum_{t = 1}^k |v^t| \cdot \left| \sum_{j \in \cl{t}} (A_{ij}-\mu_{rj} )\right| 
 \leq  \sum_{t=1}^k \frac{ \left| \sum_{j \in \cl{t}} A_{ij} - P_{rt} \cdot n_t \right| }{\sqrt{n_t}}
\end{eqnarray*}
where $n_t$ denotes the size of cluster $\cl{t}$. The last inequality above follows from the fact that $|v| = 1$. So,
 $1 \geq \sum_{j \in \cl{t}} v_j^2
= (v^t)^2 \cdot n_t$. Now, if $A_i$ does not satisfy the condition of the lemma, then  there must be some $t$ for which
\begin{eqnarray*}
 \left| \sum_{j \in \cl{t}} A_{ij} - P_{rt} \cdot n_t \right| \geq c \sigma \sqrt{n_t} \cdot \left(  \log \left( \frac{n}{\delta} \right) +
\frac{1}{\sqrt{w_{\min}}} \right)
\end{eqnarray*}
Now note that $A_{ij}$, $j \in \cl{r}$ are i.i.d. $0$-$1$ random variables with mean $P_{rt}$.
Now we use the following version of Chernoff bound : let $X_1, \ldots, X_l$ be i.i.d. 0--1 random variables, each with mean $p$. Then
$$\Pr \left[ \left| \sum_{i = 1}^l X_i - l \cdot p \right| \geq \eta
 \cdot l p \right] \leq \left\{ \begin{array}{ll} e^{-\eta^2 l p/4} & \mbox{if $\eta \leq 2e-1$ } \\
2^{-\eta \cdot l p } & \mbox{ otherwise } \end{array} \right. , $$
For us, $\eta = \frac{c \sigma}{P_{rt} \sqrt{n_t}} \cdot \left(  \log \left( \frac{n}{\delta} \right) + \frac{1}{\sqrt{w_{\min}}} \right). $
 If $\eta \leq 2e-1$, the probability of this event is at most $$\exp \left( - \frac{c^2 \sigma^2}{n_t P_{rt}^2} \cdot n_t P_{rt} \cdot \log \left( \frac{n}{\delta} \right) \right)  \leq \frac{\delta}{n^2}. $$
Now, assume $\eta > 2e -1$. In this case the probability of this event is at most
$$2^{-c  \sigma \sqrt{ \frac{n_t}{w_{\min}} } } \leq
\frac{1}{n^3}, $$
where we have assumed that $\sigma \geq \frac{3 \log n }{n}$ (we need this assumption anyway to use Wigner's theorem for bounding $||A-C||$).
}

Assuming $$ | \mu_r - \mu_s| \geq  4 ck \cdot \sigma \cdot \left( \log \left( \frac{n}{\delta} \right)
+ \frac{1}{\sqrt{w_{\min}}} \right), $$ we see that
$$|{\hat {A_i}} - \mu_s| - |{\hat {A_i}} - \mu_r| \geq  \frac{c k ||A-C||}{\sqrt{n_r}} + \frac{c k ||A-C||}{\sqrt{n_s}}, $$
with probability at least $1 - \frac{\delta}{nk}$.
Here,  we have used the fact that  $||A-C|| \leq c' \cdot \sigma  \sqrt{n}$ with high probability (Wigner's theorem).
Now, using union bound, we get that all the points satisfy the proximity condition with probability at least $1- \delta$.

\noindent
{\bf Remark : } Here we have used $C$ as the matrix whose rows are the actual means $\mu_r$. But while applying Theorem~\ref{thm:main}, $C$ should represent
 the means of the samples in $A$ belonging to a particular cluster. The error incurred here can be made very small and will not affect the results.
So we shall assume that $\mu_r$ is the actual mean of points in $T_r$. Similar comments apply in other applications described next.
\subsection{Learning Mixture of Gaussians}
We are given a mixture of $k$ Gaussians $F_1, \ldots, F_k$ in $d$ dimensions. Let the mixture weights of these distributions be
$w_1, \ldots, w_k$ and $\mu_1, \ldots, \mu_k$ denote their means  respectively.

\begin{lemma}
Suppose we are given a set of $n = \poly \left( \frac{d}{w_{\min}} \right)$ samples from the mixture distribution. Then these points
satisfy the proximity condition with high probability if $$|\mu_r - \mu_s| \geq \frac{c k \sigma_{\max}}{\sqrt{w_{\min}}} \polylog \left( \frac{d}{w_{\min}}
\right), $$
for all $r, s, r \neq s$. Here $\sigma_{\max}$ is the maximum variance in any direction of any of the distributions $F_r$.
\end{lemma}

\proof{ It can be shown that $||A-C||$ is $O \left( \sigma_{\max} \sqrt{n} \cdot \polylog \left( \frac{d}{w_{\min}} \right) \right) $
with high probability (see \cite{DasguptaHKM07}).
Further, let $p$ be a point drawn from the distribution $F_r$. Let $L_{rs}$ be the line joining $\mu_r$ and $\mu_s$. Let $\hat p$ be the projection
of $p$ on this line. Then the fact that $F_r$ is Gaussian implies that $|{\hat p} - \mu_r| \leq \sigma_{\max} \polylog(n)$ with probability at least
$1 - \frac{1}{n^2}$. It is also easy to check that the number of points from $F_r$ in the sample is close to $w_r n$ with high probability.
Thus, it follows that all the points satisfy the proximity condition with high probability.
}

The above lemma and Theorem~\ref{thm:main} imply that we can correctly classify all the points. Since we shall sample
at least $\poly(d)$ points from each distribution,
we can learn each of the distribution to high accuracy.

%These imply results from \cite{AchlioptasM05} and \cite{KannanSV08}, except those results only require a separation depending on standard deviations of particul
\subsection{Learning Mixture of  Distributions with Bounded Variance}
\label{sec:boundedvariance}
We consider a mixture of distributions $F_1, \ldots, F_k$ with weights $w_1, \ldots, w_k$. Let $\sigma$ be an upper bound on the variance along
any direction of a point sampled from one of these distributions. In other words,
$$ \sigma \geq E \left[ \left( (x-\mu_r) \cdot v \right)^2 \right], $$
for all distributions $F_r$ and each unit vector $v$. %Here the expectation is over a sample $x$ drawn from $F_r$.

\begin{theorem}
Suppose we are given a set of $n = \poly \left( \frac{d}{w_{\min}} \right)$ samples from the mixture distribution. Assume that
$\sigma \geq \frac{\polylog(n)}{\sqrt{d}}$. Then there is an algorithm to correctly classify at least $1 - \eps$ fraction of the points
provided
 $$|\mu_r - \mu_s| \geq \frac{40 k \sigma}{\sqrt{\eps}} \polylog \left( \frac{d}{\eps}
\right), $$
for all $r, s, r \neq s$. Here  $\eps$ is assumed to be less than $w_{\min}$.
\end{theorem}

\proof{ The algorithm is described in Figure~\ref{fig:alg1}. We now prove that this algorithm has the desired properties. Let $A$
denote the $n \times d$ matrix of points and $C$ be the corresponding matrix of means. We first bound the spectral norm of $A-C$. The
bound obtained is quite high, but is probably tight.

\begin{figure}[htb]
  \hrule
  \begin{tabularx}{\textwidth}{|X|}
\begin{enumerate}
    \item Run the first step of Algorithm {\tt Cluster} on the set of points, and let $\nu_1, \ldots, \nu_k$ denote the centers obtained.
    \item Remove centers $\nu_r$ (and points associated with them)
 to which less than $d^2 \log d$ points are assigned. Let $\nu_1, \ldots, \nu_{k'}$ be the remaining centers.
    \item Remove any point whose distance from the nearest center in $\nu_1, \ldots, \nu_{k'}$ is more than $\frac{\sigma \sqrt{n}}{\sqrt{d}}$.
    \item Run the algorithm {\tt Cluster} on the remaining set of points and output the clustering obtained.
\end{enumerate}
  \end{tabularx}
  \hrule
  \caption{Algorithm for Clustering points from mixture of distributions with bounded variance. }
  \label{fig:alg1}
\end{figure}

\begin{lemma}
\label{lem:highbound}
With high probability, $||A-C|| \leq \sigma \sqrt{dn} \cdot \polylog(n). $
\end{lemma}

\proof{ We use Fact~\ref{fact:nice}. Let $Y$ denote $A-C$. Note that $|Y_i|^2 = \sum_{j=1}^d (A_{ij} - C_{ij})^2 $.
Since $E[(A_{ij} - C_{ij})^2] \leq \sigma^2$ (because it is the variance of this distribution along one of the coordinate axes),
the expected value of $|Y_i|^2$ is at most $\sigma^2 d$. Now using Chebychev's inequality, we see that $\max_i |Y_i| \geq \sigma \sqrt{dn} \polylog(n)$
is at most $\frac{1}{\polylog (n)}$. Now we consider $Y^TY = \sum_i E[Y_i^T Y_i]$ (recall that we are treating $Y_i$ as a row vector).
So if $v$ is a unit (column) vector, then $v^T E[Y^TY] v = \sum_i E[|Y_i \cdot v|^2]$. But $E[|Y_i \cdot v|^2]$ is just the
the variance of the distribution corresponding to $A_i$ along $v$. So this quantity is at most $\sigma^2$ for all $v$. Thus, we
see that $||E[Y^TY|| \leq \sigma^2 n$. This proves the lemma.
}

The above lemma allows us to bound the distance between $\mu_r$ and the nearest mean obtained in Step 1 of the algorithm. The proof proceeds
along the same lines as that of Lemma~\ref{lem:case0}.
\begin{lemma}
\label{lem:distmeanest}
For each $\mu_r$, there exits a center $\nu_r$ such that $\nu_r$ is not removed in Step 2 and
$|\mu_r - \nu_r| \leq \frac{10 \sigma \sqrt{dk}}{\sqrt{\eps}} \cdot \polylog(n)$.
\end{lemma}
\proof{ Suppose the statement of the lemma is false for $\mu_r$. At most $k d^2 \log d $, which is much less than $|T_r|$, points are assigned to
a center which is removed in Step 2. The remaining points in $T_r$ are assigned to centers which are not removed.
 So, arguing as in the proof of Lemma~\ref{lem:case0}, the clustering cost in step 1 for points in $T_r$ is at least
\begin{eqnarray*}
 \frac{|T_r| - kd^2 \log d}{2} \cdot \left(  \frac{10 \sigma \sqrt{dk}}{\sqrt{\eps}} \cdot \polylog(n) \right)^2 - \sum_{i \in T_r} (\mu_r - {\hat A}_i)^2
& \geq & 50 \sigma^2 dk n \cdot \polylog(n) - ||A-C||^2 \\
& \geq & 10k ||A-C||^2
\end{eqnarray*}

where the last inequality follows from Lemma~\ref{lem:highbound}. But, as in the proof of Lemma~\ref{lem:case0}, this is a contradiction.
}

Note that in the lemma above, $\nu_r$ may not be unique for different means $\mu_r$. Call a point $A_i \in T_r$ {\em bad} if $|A_i - \mu_r|
\geq  \frac{\sigma \sqrt{n}}{2}$.  Call a point $A_i \in T_r$ {\em nice} if $|A_i - \mu_r|
\leq  \frac{\sigma \sqrt{n}}{2 \sqrt{d}}$
\begin{lemma}
\label{lem:bad}
The number of bad points is at most $d \cdot \log d$ with high probability. The number of points which are not nice is at most $d^2 \log d$
with high probability. The number of nice points that are removed is at most $4 kd^2 \log d$.
\end{lemma}
\proof{ Arguing as in the proof of Lemma~\ref{lem:highbound}, the probability that $|A_i - C_i| \geq \sigma \cdot \sqrt{n}$ is at most
$\frac{d}{n}$. So the expected number of bad points is at most $d$. The first statement in the lemma now follows from Chernoff bound.
The second statement is proved similarly. At most $kd^2 \log d$ points are removed in Step 1. Now suppose $A_i$ is nice. Then Lemma~\ref{lem:distmeanest}
implies that it will not be removed in Step 3 (using Lemma~\ref{lem:highbound} and the fact that $n$ is large enough).
}

\begin{corollary}
\label{cor:bad}
With high probability the following event happens : suppose $\nu_r$ does not get removed in Step 2. Then there is a mean $\mu_r$ such that
$|\mu_r - \nu_r| \leq \frac{2 \sigma \sqrt{n}}{\sqrt{d}}. $
\end{corollary}
\proof{ Since $\nu_r$ is not removed, it  has at least one nice point $A_i$ assigned to it (otherwise it will have at most $d^2 \log d$ points
assigned to it and it will be removed). The distance of $A_i$ to the nearest mean $\mu_s$ is at most $\frac{\sigma \sqrt{n}}{2 \sqrt{d}}$,
and Lemma~\ref{lem:distmeanest} implies that there is a center $\nu_s$ which is not removed and for which $|\mu_s - \nu_s|
\leq \frac{\sigma \sqrt{n}}{2 \sqrt{d}}$. So, $|\nu_s - A_i| \leq \frac{ \sigma \sqrt{n}}{\sqrt{d}}$.
Since $\nu_r$ is the closest center to $A_i$, $|\nu_r - A_i| \leq \frac{ \sigma \sqrt{n}}{\sqrt{d}}$ as well. Now, $|\nu_r - \mu_s| \leq
|\nu_r - A_i| + |A_i - \mu_s|$ and the result follows.
}

Let $A'$ be the set of points which are remaining after the third
step of our algorithm.
 We now define a new clustering $T_1', \ldots, T_k'$ of points in $A'$. This clustering will be very close to the
actual clustering $T_1, \ldots, T_k$ and so it will be enough to correctly cluster a large fraction of the points according to this new
clustering. We define
\begin{eqnarray*}
T_r' &  = &  \{ A_i \in T_r : \mbox{$A_i$ is not bad and does not get removed } \} \cup \{A_i : \mbox{$A_i$ is a bad point which } \\
  & & \hspace*{ 0.8 in} \mbox{
 does not get removed and the nearest center among the actual centers is $\mu_r$} \}.
\end{eqnarray*}
%$T_r'$ contains all the points in $T_r$ which are not bad and do not get removed.
%Now consider a bad point $p$ which does not get removed. Let $\mu_r$ be the closest mean to $p$. Assign $p$ to $T_r'$.
 Let $\mu_r'$ be the mean of $T_r'$ and $C'$ be the
corresponding matrix of means.
\begin{lemma}
With high probability, for all $r$, $||A'-C'|| \leq O(\sigma \cdot \sqrt{n} \cdot \polylog(n))$, and $|\mu_r - \mu_{r'}| \leq
\frac{10 \sigma k d^2  \log d }{\eps \sqrt{n}}$.
\end{lemma}
\proof{ We first prove the second statement. The  points in $T_r'$ contain all the points in $T_r$ except for at most $5 kd^2  \log d$
points (Lemma~\ref{lem:bad}). First consider the points in $T_r - T_r'$ which are not bad.
 Since all these points are at distance less than $\sigma \sqrt{n}$ from $\mu_r$, the removal of these points shifts the
mean by at most $\frac{5 \sigma k \sqrt{n} d^2  \log d}{|T_r|} \leq \frac{ \sigma k d^2 \log d }{\eps \sqrt{n}}$. Now $T_r'$ may contain some
bad points as well. First observe that any such bad point must be at most $\frac{3 \sigma \sqrt{n}}{\sqrt{d}}$ away from $\mu_r$. Indeed, the reason
why we retained this bad point in Steps 2 and 3 is because it is at distance at most $\frac{\sigma \sqrt{n}}{\sqrt{d}}$ from $\nu_r$ from some $r$.
So combined with Corollary~\ref{cor:bad}, this statement is true.
 So these bad points can again shift the mean by a similar amount. This proves the second
part of the lemma.

Now we prove the first part of the lemma. Break $A'-C'$ into two parts -- $A_B'-C_B'$ and $A_G'-C_G'$ -- the rows of $A'-C'$ which are from bad points
 and the
remaining rows (good) respectively.
$A_B'-C_B'$ has at most $d \log d$ rows, and each row (as argued above) has length at most $\frac{4 \sigma \sqrt{n}}{\sqrt{d}}$.
So $||A_B'-C_B'|| \leq 3 \sigma \sqrt{n} \log(d)$. Now consider $A_G'-C_G'$. Let $C_G$ be the rows of the original matrix $C$ corresponding to
to $A_G'$. Then $||A_G'-C_G'|| \leq ||A_G'-C_G|| + ||C_G'-C_G||$. Each row of $C_G-C_G'$ has length at most
 $\frac{10 \sigma k d^2  \log d }{\eps \sqrt{n}}$ and so its spectral norm is at most $\frac{10 \sigma k d^2  \log d }{\eps}$, which is quite small
(doesn't involve $n$ at all). So it remains to bound $||A_G'-C_G||$. Let $Z$ be the rows of $A-C$ which correspond to points which are not
bad. Note that the rows of $Z$ are independent and have length at most $\sigma \sqrt{n}$.  So applying Fact~\ref{fact:nice} and arguing as in Lemma~\ref{lem:highbound},
we can show that $||Z||$ is at most $\sigma \sqrt{n} \cdot \polylog(n)$. Now observe that $A_G'-C_G$ is obtained by picking some rows
of $Z$ (by a random process), and so its spectral norm is at most that of $||Z||$.  This proves the lemma.
}

We are now ready to prove the main theorem. We would like to recover the clustering $C'$ (since $C$ and $C'$ agree on all but the bad points).
 We argue that at least $(1-\eps)$ fraction of the
points satisfy the proximity condition. Indeed, it is easy to check that at least $(1-\eps)$ fraction of the points $A_i$ are at distance at most
$\frac{4 \sigma \sqrt{d}}{\sqrt{\eps}}$ from the corresponding mean $\mu_r$ and satisfy the proximity condition. Since the distance between
$\mu_s$ and $\mu_s'$ is {\em very small} (dependent inversely on $n$), and $A_i$ is only $\frac{4 \sigma \sqrt{d}}{\sqrt{\eps}}$ far from
$\mu_r$, it will satisfy the proximity condition for $A', C'$ as well (provided $n$ is large enough).
}
\subsection{Sufficient conditions for convergence of $k-$means}\label{sec:convergence}

As mentioned in Section~\ref{sec:previous}, Ostrovsky et. al. \cite{OstrovskyRSS06} provided the first sufficient conditions under which they prove effectiveness of (a suitable variant  of) the $k-$means algorithm.
Here, we show that their conditions are (much) stronger than the proximity condition. We first describe their conditions.
Let
$\Delta_k$  be the optimal cost of the $k$-means problem (i.e.,  sum of distance squared
  of each point to nearest center) with $k$ centers.  They  require:
$$\Delta_k\leq \eps \Delta_{k-1}.$$

\begin{claim}
The condition above implies the proximity condition for all but $\eps$ fraction of the points.
\end{claim}

\proof{ Suppose the above condition is true. One way of getting a solution with $k-1$ centers is to remove a center
$\mu_r$ and move all points in $T_r$ to the nearest other center $\mu_s$. Now, their condition implies
$$|\mu_r-\mu_s|^2\geq \frac{1}{n_r \cdot \eps}||A-C||_F^2\quad\forall s\not= r.$$
 If some $\eps$ fraction of $T_r$ do not satisfy the proximity condition, then
the distance squared of each such point to $\mu_r$ is at least the distance squared
along the line $\mu_r$ to $\mu_s$ which is at least $(1/4)|\mu_r-\mu_s|^2$ which is
at least $\Omega (||A-C||_F^2/n_r\epsilon)$. So even the assignment cost of such points exceeds $||A-C||_F^2$, the total cost, a contradiction.
This proves the claim.
}

%\cite{OstrovskyRSS06}'s aim is a PTAS, rather than actual convergence. They show that under their condition, just one iteration of $k-$means suffices.
We now show that our algorithm gives a PTAS for the $k-$means problem.

\paragraph{Getting a PTAS}
Let $T_1, \ldots, T_k$ be the optimal clustering and $\mu_1, \ldots, \mu_k$ be the corresponding means. As before, $n_r$ denotes the
size of $T_r$. Let $G$ be the set of points which satisfy the proximity condition (the good points). The above claim shows that
$|G| \geq (1-\eps)n$. %But we can refine the above claim to a stronger statement as follows.
%Let $\Delta_{rs}'$ denote
%$\frac{||A-C||}{2 \sqrt{\eps} \cdot \sqrt{\min(n_r, n_s)}}. $ Note that $\Delta_{rs} \leq \Delta_{rs}'$ (assuming $\eps$ is small enough).
%Then we can easily modify the above claim to prove that at most $\eps$ fraction of the points do not satisfy the proximity
%condition with respect to $\Delta_{rs}'$.
 For simplicity, assume that {\em exactly} $\eps$ fraction of the points do not satisfy the
proximity condition.

Let $S_1, \ldots, S_r$ be the
clustering output by our algorithm. % Let $M$ denote the set of mis-classified good points, i.e., points in $\cup_r |G \cap (T_r - S_r)|$.
%Our main theorem states that $|M|$ is $O(k^2 \eps n)$.
Let $\mu_r'$ be the mean of $S_r$. First observe that Theorem~\ref{thm:key} implies
that when our algorithm stops,
\begin{eqnarray}
\label{eqn:bounds}
|\mu_r - \mu_r'| \leq \frac{c \cdot \sqrt{k \eps n}}{n_r} \cdot ||A-C||.
\end{eqnarray}
for some constant $c$.
 For a point $A_i$, let $\alpha(A_i)$ the
square of its distance to the closest mean among $\mu_1, \ldots, \mu_k$. Define $\alpha'(A_i)$ for the solution output by our algorithm similarly.

\begin{claim}
\label{cl:part1}
If $A_i \notin G$, then $\alpha'(A_i) \leq (1 + O(\eps)) \cdot \alpha(A_i). $
\end{claim}
\proof{ Suppose $A_i \in T_r$, and it does not satisfy the proximity condition for the pair $\mu_r, \mu_s$. Then it is easy to see that $\alpha(A_i)
\geq \left( |\mu_r - \mu_s|/2 - \Delta_{rs} \right)^2 \geq \frac{|\mu_r - \mu_s|^2}{4} \geq \frac{||A-C||^2}{4 n_r \eps }$.
Let $\mu_t$ be the closest mean to $A_i$. Then
\begin{eqnarray*}
\alpha'(A_i) & \leq & |A_i - \mu_t'|^2 \ \leq \ (1+\eps) |A_i - \mu_t|^2 + \left( \frac{1}{\eps} + 1 \right)  \cdot |\mu_t - \mu_t'|^2 \\
& \leq & (1+\eps) \alpha(A_i) + \frac{c' k n}{n_r^2} \cdot ||A-C||^2  \leq  (1 + O(\eps)) \alpha(A_i)
\end{eqnarray*}
where the second last inequality follows from equation~(\ref{eqn:bounds}). Note that the  constant in $O(\eps)$ above contains terms involving $k$ and $w_{\min}$.
}

\begin{claim}
\label{cl:part2}
If $A_i \in G$, but is mis-classified by our algorithm, then $\alpha'(A_i) \leq (1+O(\eps)) \cdot \alpha(A_i). $
\end{claim}

\proof{ Suppose $A_i \in G \cap T_r \cap S_s$. We use the machinery developed in the proof of Theorem~\ref{thm:iterate}. Define $\lambda, u$
as in the proof of this theorem. Clearly, $\alpha'(A_i) \leq \alpha(A_i) + 2 |\mu_r - \mu_s|^2$ (here we have also used equation~(\ref{eqn:bounds})).
But note that $\alpha(A_i) \geq |u|^2$. Now, $|u| \geq \frac{\Delta_{rs} |\mu_r-\mu_s|}{64 \delta} = \Omega \left( \frac{|\mu_r-\mu_s|}{\sqrt{\eps}} \right)$
(again using equation~(\ref{eqn:bounds})). This implies the result.
}

\begin{claim}
\label{cl:part3}
For all $r$,
$$\sum_{A_i \in G \cap T_r \cap S_r} \alpha'(A_i) \leq (1+O(\sqrt{\eps})) \cdot \sum_{A_i \in G \cap T_r \cap S_r} \alpha(A_i) +
O(\sqrt{\eps}) \cdot \sum_{A_i \notin G} \alpha(A_i) . $$
\end{claim}
\proof{ Clearly, $$|A_i - \mu_r'|^2 \leq (1+\sqrt{\eps} \cdot \beta) |A_i -\mu_r|^2 + \left( 1 + \frac{1}{\sqrt{\eps} \cdot \beta} \right) |\mu_r - \mu_r'|^2, $$
where $\beta$ is a large constant in terms of $k, \frac{1}{w_{\min}}, c$. Summing this over all points in $T_r \cap S_r$, we get
\begin{eqnarray*}
\sum_{A_i \in G \cap T_r \cap S_r} \alpha'(A_i) \leq (1 + O(\sqrt{\eps})) \sum_{A_i \in G \cap T_r \cap S_r} \alpha(A_i) + \frac{\sqrt{\eps} ||A-C||^2}{4k}
\end{eqnarray*}
where the last inequality follows from (\ref{eqn:bounds}) assuming $\beta$ is large enough. But now, the proof of Claim~\ref{cl:part1}, implies
that $\sum_{A_i \notin G} \alpha(A_i) \geq \frac{||A-C||^2}{4}. $ So we are done.
}
Now summing over all $r$ in Claim~\ref{cl:part3} and using Claims~\ref{cl:part1}, \ref{cl:part2} implies that our algorithm is also a PTAS.
\section{Boosting}\label{boosting}
Recall that the proximity condition requires that the distance between the means be polynomially dependent on $\frac{1}{w_{\min}}$ -- this
could be quite poor when one of the clusters is considerably smaller that the others. In this section, we try to overcome this obstacle for a
 special class of distributions.

Let $F_1, \ldots, F_k$ be a mixture of distributions in $d$ dimensions. Let $A$ be the $n \times d$ matrix of samples from the distribution and $C$
be the corresponding matrix of centers. Let $D_{\min}$ denote $\min_{r, s, r \neq s} |\mu_r - \mu_s|$.
 Then the key property that we desire from the mixture of distributions is as follows. The following
conditions  should be satisfied with high probability :
\begin{enumerate}
\item For all $r, s$, $r \neq s$,
\begin{eqnarray}
\label{eqn:condn0}
|\mu_r - \mu_s| \geq \frac{10 k ||A-C||}{\sqrt{n}}
\end{eqnarray}
\item
For all $i$,
 \begin{eqnarray}
\label{eqn:condn1}
%|A_i - C_i| \leq \frac{||A-C||}{\sqrt{n}} \cdot \sqrt{d} n^{\alpha} \polylog(n),
|A_i - C_i| \leq D_{\min} \cdot \sqrt{d} n^{\alpha} \polylog(n),
\end{eqnarray}
where $\alpha$ is a small enough constant (something like 0.1 will suffice).
\item For all $r, s, r \neq s$,
\begin{eqnarray}
\label{eqn:condn2}
\sum_{i \in T_r} \left[(A_i - \mu_r) \cdot v \right]^2 \leq \frac{|\mu_r - \mu_s|^2}{16} \cdot |T_r|
\end{eqnarray}
where $v$ is the unit vector joining $\mu_r$ and $\mu_s$. This condition is essentially saying that the average variance of points in $T_r$
along $v$ is bounded by $\frac{|\mu_r - \mu_s|}{\sqrt{4}}$.
\end{enumerate}
The number of samples $n$ will be a polynomial in $\frac{d}{w_{\min}}$.
Recall that  $D_{\min}$ denotes $\min_{r,s, r \neq s} |\mu_r  \mu_s|$.
To simplify the presentation, we assume that $|\mu_r - \mu_s| \leq D_{\min} \cdot \left(\frac{d}{w_{\min}} \right)^{\beta}$
for a constant $\beta$ for all
pairs $r, s$.  We will later show how to get rid of this assumption.  We now sample two sets of $n$ points from this distribtion  -- call these $A$ and $B$.
  Assume that both $A$ and $B$ satisfy the
condtions (\ref{eqn:condn1}) and (\ref{eqn:condn2}).
For all $r$, we assume that the mean of $A_i, i \in T_r$ is $\mu_r$  and $T_r \cap A$ has size $w_r \cdot n$.
We assume the same for the points in $B$. The error caused by removing this
assumption will not change our results. Let $\mu$ denote the overall mean of the points in $A$ (or $B$).
 Note that $\mu = \sum_r w_r \mu_r $.
We  translate the points so that the overall mean is 0. In other words, define a translation $f$ as $f(x) = x - \mu$. Let $A_i'$ denote
$f(A_i)$. Define $B_i'$ similarly. We now define a set $X$ of $n$ points in $n$ dimensions. The point $X_i$ is defined as
$$\left(A_i' \cdot B_1', \ldots, A_i' \cdot B_n' \right). $$
The correspondence between $X_i$ and $A_i$ naturally defines a partitioning of $X$ into $k$ clusters. Let $S_r, r =1 , \ldots, k,$ denote
these clusters. The mean $\theta_r$ of $S_r$ is  $$\left(C_r' \cdot B_1', \ldots, C_r' \cdot B_n' \right), $$
where $C_r' = C_r - \mu$.
 Let $Z$ denote the matrix of means of $X$, i.e., $Z_i = \theta_r$ if $X_i \in S_r$. We now show that this process {\em amplifies} the distance
between the means $\theta_r$ by a much bigger factor than $\frac{||X-Z||}{\sqrt{n}}$.

\begin{lemma}
\label{lem:distmeans}
For all $r, s, r \neq s$,
$$ |\theta_r - \theta_s| \geq \frac{|\mu_r - \mu_s|^2 }{4} \cdot \sqrt{w_{\min}} \sqrt{n}. $$
\end{lemma}
\proof{ First observe that $(\mu_r - \mu_s) \cdot (\mu_r-\mu_s) = (\mu_r - \mu) \cdot (\mu_r - \mu_s) -
(\mu_s - \mu) \cdot (\mu_r-\mu_s)$. So at least one of $|(\mu_r - \mu) \cdot (\mu_r-\mu_s)|, |(\mu_s - \mu)\cdot (\mu_r-\mu_s)|$ must be
at least $\frac{|\mu_r - \mu_s|^2}{2}$. Assume without loss of generality that this is so for $|(\mu_r - \mu)\cdot(\mu_r-\mu_s)|$. Now consider the coordinates $i$ of
$\theta_r - \theta_s$ corresponding to $S_r$. Such a coordinate will have value $(\mu_r - \mu_s) \cdot B_i'$. Therefore,
\begin{eqnarray*}
|\theta_r - \theta_s|^2 & \geq & \sum_{i \in S_r} \left[(\mu_r - \mu_s) \cdot (B_i-\mu) \right]^2 \\
& \geq &  \frac{|S_r|}{2} \left[ (\mu_r - \mu_s) \cdot (\mu-\mu_r)\right]^2 - \sum_{i \in S_r}  \left[(\mu_r - \mu_s) \cdot (B_i-\mu_r) \right]^2 \\
& \geq & \frac{|S_r|}{16} \cdot |\mu_r - \mu_s|^4
\end{eqnarray*}
where the last inequality follows from (\ref{eqn:condn2}). This proves the lemma.
}

Now we bound $||X-Z||$.
\begin{lemma}
\label{lem:normbound}
With high probability, $$\frac{||X-Z||}{\sqrt{n}} \leq
  D_{\min}^2 \cdot d \cdot n^{2\alpha} \cdot \left(\frac{d}{w_{\min}} \right)^{\beta}
\cdot \polylog(n) $$
\end{lemma}
\proof{ Let $Y$ denote the matrix $X-Z$, and $D$ denote the matrix $E[Y^TY]$ where the expectation is over the choice of
$A$ and $B$. We shall use Fact~\ref{fact:nice} to bound $||Y||$.
Thus, we just need to bound $\max_i |Y_i|$ and $||D||$. Let $\gamma$ denote $ D_{\min}^2
 \cdot d \cdot n^{2\alpha}
\cdot \left(\frac{d}{w_{\min}} \right)^{\beta}
\cdot \polylog(n) $.
\begin{claim}
\label{cl:bound1}
For all $i$, $$|Y_i| \leq \sqrt{n} \cdot \gamma $$
\end{claim}
\proof{ Suppose $X_i \in S_r$. Then the $j^{th}$ coordinate of $Y$ is $(A_i - \mu_r) \cdot (B_j-\mu) = (A_i - \mu_r) \cdot \left((B_j - \mu_{r'})
+ (\mu - \mu_{r'}) \right)$, where $r'$ is such that $B_j \in T_{r'}$. Now, condition~(\ref{eqn:condn1}) implies
that $|A_i-\mu_r|, |B_j - \mu_{r'}| \leq D_{\min} \cdot \sqrt{d} n^{\alpha} \polylog(n)$. This implies the claim.
}
\begin{claim}
\label{cl:bound2}
$$||D|| \leq \gamma^2 n $$
\end{claim}
\proof{ We can write $Y^TY$ as $\sum_i Y_i^T Y_i$. Let $v$ be any unit vector. Then $v^T E[Y^TY] v = \sum_i E|Y_i \cdot v|^2$.
For a fixed $i$, where $A_i \in T_r$,
\begin{eqnarray*}
E|Y_i \cdot v|^2 & = & E \left( \sum_j v_j \left[ (X_i - \mu_r) \cdot (Y_j - \mu) \right] \right)^2\\
& = &  \sum_j v_j^2 E \left[ (X_i - \mu_r) \cdot (Y_j - \mu) \right]^2
\end{eqnarray*}
where the last inequality follows from the fact that expectation of $Y_j$ is $\mu$ and $Y_j, Y_j'$ are independent if $j \neq j'$.
Rest of the argument is as in Claim~\ref{cl:bound1}.
}
The above two claims combined with Fact~\ref{fact:nice} imply the lemma.
}

Now we pick $n$ to be at least $\left( \frac{d}{w_{\min}} \right)^{8 (\beta + 1)}$. Assuming $\alpha < 0.1$, this implies (using the above
two lemmas) that for all
$r,s$, $r \neq s$,
\begin{eqnarray}
\label{eqn:gap}
|\theta_r - \theta_s| \geq \frac{||X-Z||}{\sqrt{n}} \cdot \left( \frac{d}{w_{\min}} \right)^{4 \beta}
\end{eqnarray}

We now run the first step of the algorithm {\tt Cluster} on $X$. We claim that the clustering obtained after the first step has very few
classification errors. Let $\phi_r, r= 1, \ldots, k$ be the $k$ centers output by the first step of the algorithm {\tt Cluster}.
Lemma~\ref{lem:case0} implies that for each center $\theta_r$, there exists a center $\phi_r$ satisfying
$$|\theta_r - \phi_r| \leq 20 \sqrt{k} \cdot \frac{ ||X-Z||}{\sqrt{w_{\min} \cdot n}}.$$ Order the centers $\phi_r$ such that $\phi_r$
is closest to $\theta_r$ -- equation~(\ref{eqn:gap}) implies that the closest estimated centers to different $\theta_r$ are distinct.
It also follows that for $r \neq s$
\begin{eqnarray}
\label{eqn:gapapprox}
|\phi_r - \phi_s| \geq \frac{1}{2} \cdot \frac{||X-Z||}{\sqrt{n}} \cdot \left( \frac{d}{w_{\min}} \right)^{4 \beta}
\end{eqnarray}

\begin{lemma}
\label{lem:error}
The number of points in $S_r$ which are not assigned to $\phi_r$ after the first step of the algorithm is at most $\left( \frac{w_{\min}}{d} \right)^{2 \beta}
\cdot n$.
\end{lemma}
\proof{ We use the notation in Step 1 of algorithm {\tt Cluster}. Suppose the statement of the lemma is not true. Then, in the $k$-means solution,
at least $\left( \frac{w_{\min}}{d} \right)^{2 \beta}
\cdot n$ points in ${\hat X}_i, i \in S_r$ are assigned to a center at least
$\frac{1}{2} \cdot \frac{||X-Z||}{\sqrt{n}} \cdot \left( \frac{d}{w_{\min}} \right)^{4 \beta}$ distance away (using equation~\ref{eqn:gapapprox}).
But then the square of $k$-means clustering cost is much larger  that $k \cdot ||X-Z||^2$.
}

Now, we use the clustering given by the centers $\phi_r$ to partition the original set of points $A$ -- thus we have a clsutering of these
points where the number of {\em mis-classified} points from any cluster $T_r$ is at most $\left( \frac{w_{\min}}{d} \right)^{2 \beta}
\cdot n$. Let $S_r$ denote this clustering, where $S_r$ corresponds to $T_r$. Let $\nu_r$ denote the center of $S_r$. We now argue that
$|\nu_r - \mu_r|$ is very small.

\begin{lemma}
\label{lem:distance}
For every $s$, $|\nu_s - \mu_s| \leq \frac{||A-C||}{\sqrt{n}}. $
\end{lemma}
\proof{ We use arguments similar to proof of Theorem~\ref{thm:key}.  Let
$n_{rs},\mu_{rs} $ denote the number and mean respectively of $T_r\cap S_s$.
 Similarly, define $n_{ss}$ and $\mu_{ss}$ as the size and mean of the points in $\cl{s} \cap S_s$.
We know that
$$|S_s|\nu_s = n_{ss}\mu_{ss}+\sum_{r \neq s} n_{rs}\mu_{rs}.$$
Theorem~\ref{thm:iterate} implies that $$|\mu_{rs} - \mu_s| \leq \frac{100 \cdot ||A-C||}{\sqrt{n_{rs}}}, $$
and Corollary~\ref{cor:move} implies that $$|\mu_{ss} - \mu_s| \leq \frac{\sqrt{|S_s|-n_{ss}}}{n_{ss}} \cdot ||A-C||. $$
Now, proceeding as in the proof of Theorem~\ref{thm:key}, we get
\begin{eqnarray*}
|\mu_s - \nu_s| & \leq &  \frac{n_{ss}}{|S_s|}|\mu_{ss}-\mu_s| +\sum_{r\neq s}\frac{n_{rs}}{|S_s|}|\mu_{rs}-\mu_s| \\
& \leq & \frac{400 \cdot ||A-C||}{|T_s|} \cdot\left( \sum_{r\neq s} \sqrt{n_{rs}}   \right)
\end{eqnarray*}
Using Lemma~\ref{lem:error} now implies the result.
}

Starting from the centers $\nu_r$, we run the second step of algorithm {\tt Cluster}. Then, we have the analogue of Theorem~\ref{thm:main}
in this setting.
\begin{theorem}
\label{thm:mainmodified}
Suppose a mixture of distribution satisfies the conditions (\ref{eqn:condn0}--\ref{eqn:condn2}) above and at least $(1-\eps)$ fraction of sampled points satisfy the proximity
condition. Then we can correctly classify all but $O(k^2 \eps)$ fraction of the points.
\end{theorem}

We now remove the assumption that $|\mu_r - \mu_s| \leq D_{\min} \cdot \left(\frac{d}{w_{\min}} \right)^{\beta}$ --
let $\gamma$ denote the latter quantity.
%Suppose we know $\frac{||A-C||}{\sqrt{m}}$ (more on this below).
We construct a graph $G = (V, E)$ as follows : $V$ is the set
of points $A \cup B$, and we join two points by an edge if the distance between them is at most $\gamma/k$. First observe that if $i, j \in T_r$, then
they will be joined by an edge provided the following condition holds (using condition~(\ref{eqn:condn1})) :
$$ D_{\min} \cdot \sqrt{d} n^{\alpha} \polylog(n) \leq D_{\min} \cdot \left(\frac{d}{w_{\min}} \right)^{\beta},$$
and the same for $A$ replaced by $B$ above.
This would hold if $\alpha < 0.1$ (recall that $n$ is roughly $\left( \frac{d}{w_{\min}} \right)^{8 (\beta + 1)}$).
 Now consider the connected components
of this graph. In each connected component, any two vertices are joined by a path of length at most $k$ (because any two vertices from the same cluster
$T_r$ have an edge between them). So the distance between any two vertices from the same component is at most $\gamma$. Therefore the distance
from the mean of two clusters in the same component is at most $\gamma$. Now, we can apply the arguments of this section to each component
independently. This would, however, require us to know the number of clusters in each component of this graph. If we treat $k$ as a constant, this
is only constant number of choices. A better way is to modify the definition of $X$ as follows : consider a point $A_i$.  Let $\mu$ denote the mean
of the points in the same component as $A_i$ in the graph $G$. Then $X_{ij} = (A_i - \mu) \cdot (B_j - \mu)$ if $A_i, B_j$ belong to the same component
of $G$, $L$ otherwise, where $L$ is a large quantity. Now note that $\theta_r - \theta_s$ will still satisfy the statement of Lemma~\ref{lem:distmeans}, because
if they are from the same component in $G$, then it follows from the lemma, otherwise the distance between them is at least $L$. But
Lemma~\ref{lem:normbound}
continues to hold without any change, and so rest of the arguments follow as they are.

\subsection{Applications}
We now give some applications of Theorem~\ref{thm:mainmodified}.
\subsubsection{Learning Gaussian Distributions}
Suppose we are given a mixture of Gaussian distribution $F_1, \ldots, F_k$. Suppose the means satisfy the following  separation condition
for all $r,s, r \neq s$ :
$$ |\mu_r - \mu_s| \geq \Omega \left( \sigma k \cdot \log \frac{d}{w_{\min}} \right), $$
where $\sigma$ denotes the maximum variance in any direction of the Gaussian distributions. Sample a set of $n = \poly \left( \frac{d}{w_{\min}} \right)
$ points. It is easy to check using Fact~\ref{fact:nice} that $||A-C||$ is $O( \sigma \sqrt{d} \cdot \log n)$. It is also easy to check that
condition~(\ref{eqn:condn2}) is satisfied with $\alpha = 0$. Therefore, Theorem~\ref{thm:mainmodified} implies the following.
\begin{lemma}
\label{lem:gaussiansepn}
Given a mixture of $k$ Gaussians satisfying the separation condition above, we can correctly classify a set $n$ samples, where $n = \poly \left( \frac{d}{w_{\min}}
\right)$.
\end{lemma}
\subsubsection{Learning Power Law Distributions}
Consider a mixture of distributions $F_1, \ldots, F_k$ where each of the distributions $F_r$ satisfies the following condition for every unit vector $v$ :
\begin{eqnarray}
\label{eqn:heavytail}
P_{X \in F_r} \left[ |(X-\mu_r) \cdot v| > \sigma t \right] \leq \frac{1}{t^{\gamma}}
\end{eqnarray}
where $\gamma \geq 2$ is a large enough constant. Let $A$ be a set of $n$ samples from the mixture. Suppose the means satisfy the following separation condition
for every $r, s, r \neq s$ :
$$ |\mu_r - \mu_s| \geq \Omega \left( \sigma k \cdot \left( \log \frac{d}{w_{\min}} + \frac{1}{\eps^{\frac{1}{\gamma}}} \right)\right). $$
First observe that since this is a special class of distributions considered in Section~\ref{sec:boundedvariance}. So, one can again prove that $\frac{||A-C||}{\sqrt{n}}$
is $O(\sigma \cdot \sqrt{d} \cdot \polylog(n))$. This is off from condition~(\ref{eqn:condn0}) by a factor of $\sqrt{d}$. But for large enough $n$,
inequality (\ref{eqn:gap}) will continue to hold. Now let us try to bound $\max_i |A_i - C_i|$.
\begin{claim} With high probability,
$$\max_i |A_i - C_i| \leq D_{\min} \cdot \sqrt{d} \cdot n^{\frac{2}{\gamma}} \cdot \polylog(n). $$
\end{claim}
\proof{ Let $e_1, \ldots, e_d$ be orthonormal basis for the space. Then $|(A_i - C_i) \cdot e_l| \leq \sigma (nd)^{\frac{1}{\gamma}} \cdot \log(n)$ for all $i$
with high probability. So, with high probability, for all $i$,
$$|A_i - C_i| \leq D_{\min} \sqrt{d} n^{\frac{2}{\gamma}}. $$
}

Finally, we verify condition~(\ref{eqn:condn2}). Let $v$ be a vector joining $\mu_r$ and $\mu_s$. Then, $E \left[ \left( (A_i - C_i) \cdot v \right)^2 \right]
$ is $O( \sigma^2) $ provided $\gamma \geq 2$. Now summing over all $A_i \in T_r$ and taking union bound for all $k^2$ choices for $v$ proves that
condition~(\ref{eqn:condn2}) is also satisfied. It is also easy to check that at least $1-\eps$ fraction of the points satisfy the proximity
condition.
So we have
\begin{theorem}
\label{thm:powerlaw}
Given a mixture of distributions where each distribution satisfies (\ref{eqn:heavytail}), we can cluster at least $1-\eps$ fraction of the points.
\end{theorem}

\bibliographystyle{alpha}
\bibliography{cluster-k}

\end{document}

%% file: classify.pstex_t
\begin{picture}(0,0)%
\epsfig{file=classify.pstex}%
\end{picture}%
\setlength{\unitlength}{2368sp}%
\begingroup\makeatletter\ifx\SetFigFont\undefined%
\gdef\SetFigFont#1#2#3#4#5{%
  \reset@font\fontsize{#1}{#2pt}%
  \fontfamily{#3}\fontseries{#4}\fontshape{#5}%
  \selectfont}%
\fi\endgroup%
\begin{picture}(10741,6927)(893,-6155)
\put(2626,-2086){\makebox(0,0)[lb]{\smash{{\SetFigFont{12}{14.4}{\rmdefault}{\mddefault}{\updefault}{$u$}%
}}}}
\put(1051,-4261){\makebox(0,0)[lb]{\smash{{\SetFigFont{12}{14.4}{\rmdefault}{\mddefault}{\updefault}{$\mu_r$}%
}}}}
\put(11401,-4411){\makebox(0,0)[lb]{\smash{{\SetFigFont{12}{14.4}{\rmdefault}{\mddefault}{\updefault}{$\mu_s$}%
}}}}
\put(2176,-6061){\makebox(0,0)[lb]{\smash{{\SetFigFont{12}{14.4}{\rmdefault}{\mddefault}{\updefault}{$\nu_r$}%
}}}}
\put(10051,-2686){\makebox(0,0)[lb]{\smash{{\SetFigFont{12}{14.4}{\rmdefault}{\mddefault}{\updefault}{$\nu_s$}%
}}}}
\put(2926,464){\makebox(0,0)[lb]{\smash{{\SetFigFont{12}{14.4}{\rmdefault}{\mddefault}{\updefault}{$A_i$}%
}}}}
\end{picture}%

%% file: cluster-k.bbl
\begin{thebibliography}{DHKM07}

\bibitem[ADK09]{AggarwalDK09}
Ankit Aggarwal, Amit Deshpande, and Ravi Kannan.
\newblock Adaptive sampling for k-means clustering.
\newblock In {\em APPROX-RANDOM}, pages 15--28, 2009.

\bibitem[AK01]{AroraKannan01}
Sanjeev Arora and Ravi Kannan.
\newblock Learning mixtures of arbitrary gaussians.
\newblock In {\em ACM Symposium on Theory of Computing}, pages 247--257, 2001.

\bibitem[AM05]{AchlioptasM05}
Dimitris Achlioptas and Frank McSherry.
\newblock On spectral learning of mixtures of distributions.
\newblock In {\em COLT}, pages 458--469, 2005.

\bibitem[AV06]{ArthurV06}
David Arthur and Sergei Vassilvitskii.
\newblock How slow is the {\it }-means method?
\newblock In {\em Symposium on Computational Geometry}, pages 144--153, 2006.

\bibitem[AV07]{ArthurV07}
David Arthur and Sergei Vassilvitskii.
\newblock k-means++: the advantages of careful seeding.
\newblock In {\em SODA}, pages 1027--1035, 2007.

\bibitem[BV08]{BrubakerV08}
S.~Charles Brubaker and Santosh Vempala.
\newblock Isotropic pca and affine-invariant clustering.
\newblock In {\em FOCS}, pages 551--560, 2008.

\bibitem[CR08]{ChaudhuriR08a}
Kamalika Chaudhuri and Satish Rao.
\newblock Beyond gaussians: Spectral methods for learning mixtures of
  heavy-tailed distributions.
\newblock In {\em COLT}, pages 21--32, 2008.

\bibitem[Das99]{dasgupta99}
Sanjoy Dasgupta.
\newblock Learning mixtures of gaussians.
\newblock In {\em IEEE Symposium on Foundations of Computer Science}, pages
  634--644, 1999.

\bibitem[Das03]{Dasgupta03a}
Sanjoy Dasgupta.
\newblock How fast is {\it }-means?
\newblock In {\em COLT}, page 735, 2003.

\bibitem[DHKM07]{DasguptaHKM07}
Anirban Dasgupta, John~E. Hopcroft, Ravi Kannan, and Pradipta~Prometheus Mitra.
\newblock Spectral clustering with limited independence.
\newblock In {\em SODA}, pages 1036--1045, 2007.

\bibitem[DHKS05]{DasguptaHKS05}
Anirban Dasgupta, John~E. Hopcroft, Jon~M. Kleinberg, and Mark Sandler.
\newblock On learning mixtures of heavy-tailed distributions.
\newblock In {\em FOCS}, pages 491--500, 2005.

\bibitem[DLR77]{Dempster77maximumlikelihood}
A.~P. Dempster, N.~M. Laird, and D.~B. Rubin.
\newblock Maximum likelihood from incomplete data via the em algorithm.
\newblock {\em Journal of the Royal Statistical Society, Series B},
  39(1):1--38, 1977.

\bibitem[DS07]{DasguptaS07}
Sanjoy Dasgupta and Leonard~J. Schulman.
\newblock A probabilistic analysis of em for mixtures of separated, spherical
  gaussians.
\newblock {\em Journal of Machine Learning Research}, 8:203--226, 2007.

\bibitem[HPS05]{Har-PeledS05}
Sariel Har-Peled and Bardia Sadri.
\newblock How fast is the k-means method?
\newblock In {\em SODA}, pages 877--885, 2005.

\bibitem[KSS10]{KumarSS10}
Amit Kumar, Yogish Sabharwal, and Sandeep Sen.
\newblock Linear-time approximation schemes for clustering problems in any
  dimensions.
\newblock {\em J. ACM}, 57(2), 2010.

\bibitem[KSV08]{KannanSV08}
Ravindran Kannan, Hadi Salmasian, and Santosh Vempala.
\newblock The spectral method for general mixture models.
\newblock {\em SIAM J. Comput.}, 38(3):1141--1156, 2008.

\bibitem[KV09]{KannanV09}
Ravi Kannan and Santosh Vempala.
\newblock Spectral algorithms.
\newblock {\em Foundations and Trends in Theoretical Computer Science},
  4(3-4):157--288, 2009.

\bibitem[Llo82]{lloyd}
S.~Lloyd.
\newblock Least squares quantization in pcm.
\newblock {\em Information Theory, IEEE Transactions on}, 28(2):129--137, 1982.

\bibitem[McS01]{McSherry01}
Frank McSherry.
\newblock Spectral partitioning of random graphs.
\newblock In {\em FOCS}, pages 529--537, 2001.

\bibitem[ORSS06]{OstrovskyRSS06}
Rafail Ostrovsky, Yuval Rabani, Leonard~J. Schulman, and Chaitanya Swamy.
\newblock The effectiveness of lloyd-type methods for the k-means problem.
\newblock In {\em FOCS}, pages 165--176, 2006.

\bibitem[VW04]{VempalaW04}
Santosh Vempala and Grant Wang.
\newblock A spectral algorithm for learning mixture models.
\newblock {\em J. Comput. Syst. Sci.}, 68(4):841--860, 2004.

\end{thebibliography}
